\documentstyle[prl, twocolumn, aps, psfig]{revtex}

\def\be{\begin{equation}}
\def\ee{\end{equation}}
\def\ba{\begin{eqnarray}}
\def\ea{\end{eqnarray}}
\begin{document}
\draft
\preprint{}
\title{Approximate Analytical Solutions to the 
Initial Data Problem of Black Hole Binary Systems }

\author{
Pedro~Marronetti${}^{(1)}$, 
Mijan~Huq${}^{(2)}$, 
Pablo~Laguna${}^{(2)}$,
Luis~Lehner${}^{(1)}$,
Richard~A.~Matzner${}^{(1)}$,
and Deirdre~Shoemaker${}^{(2)}$}

\address{
${}^{(1)}$ Center for Relativity, The University of Texas at
Austin, TX 78712-1081\\
${}^{(2)}$ Department of Astronomy \& Astrophysics and\\
Center for Gravitational Physics \& Geometry\\
Pennsylvania State University, University Park, PA 16802
}

\date{\today}
\maketitle
\begin{abstract}
We present approximate analytical solutions to the Hamiltonian and
momentum constraint equations, corresponding to systems composed of 
two black holes with arbitrary linear and angular momentum. The 
analytical nature of these initial data solutions makes them easier 
to implement in numerical evolutions than the traditional numerical 
approach of solving the elliptic equations derived from the Einstein
constraints. Although in general the problem of setting up initial 
conditions for black hole binary simulations is complicated by the 
presence of singularities, we show that the methods presented in this 
work provide initial data with $l_1$ and $l_\infty$ norms of
violation of the constraint equations falling below those of the 
truncation error (residual error due to discretization)
present in finite difference codes for the range of grid resolutions 
currently used. Thus, these data sets are suitable for use in evolution 
codes. Detailed results are presented for the case of a head-on 
collision of two equal-mass $M$ black holes with specific angular
momentum $0.5M$ at an initial  separation of $10M$. A straightforward
superposition method yields data adequate for resolutions of $h=M/4$,
and an ``attenuated" superposition yields data usable to resolutions 
at least as fine as $h=M/8$. In addition, the attenuated approximate
data may be more tractable in a full (computational) exact solution
to the initial value problem.
\end{abstract}
\pacs{PACS number(s): 04.70.Bw,04.25.Dm}

\narrowtext

\section{Introduction}
The computation of gravitational wave production from the interaction
and merger of compact astrophysical objects is an analytical and
computational challenge. These calculations would be able to provide 
both a predictive and an analytical resource for the
gravitational-wave interferometric detectors such as LIGO \cite{LIGO},
Virgo and GEO600 \cite{VIRGO} soon to be online. We concentrate 
on the case of binary black hole mergers \cite{Group}.

Besides ours, the other major efforts to numerically simulate black 
hole coalescences are being carried out by the AEI-WashU-NCSA \cite{PWS} 
and the Cornell-Illinois collaborations \cite{CIN}.
The resolutions currently under consideration by the AEI-WashU-NCSA
effort are as fine as $h = M/5$ in a computational domain of with $385^3$
mesh-points, where $h$ denotes the grid spacing. Our computations are 
carried out at a similarly modest  resolution $h= M/4$, on $161^3$ 
domains. The goal of all these groups is to perform eventually simulations 
in domains of up to $1000^3$ grid-points, which will allow finer 
resolution and/or large physical domain sizes.

Regardless of resolution, in order to carry out such simulations, we
must construct initial data sets representing binary black hole systems.
These data sets should not only satisfy the Einstein constraints
but also carry the desired physical content. Binary black hole initial 
data sets are the subject of this paper.

Under the $3+1$ formulation of General Relativity, the construction of 
initial data requires solving the Hamiltonian and momentum constraints:
\begin{equation}
R+\frac{2}{3} K^2 - A_i^{~j} A_j^{~i}  =  0
\label{HC}
\end{equation}
and
\begin{equation}
\nabla_j A_{i}^{~j} - \frac{2}{3} \nabla_i K  =  0~~,
\label{MC}
\end{equation}
respectively. 
Above, $R$ is the 3-dimensional Ricci scalar constructed from the
spatial
3-metric $g_{ij}$, and
\be
K_{ij} = A_{ij} +\frac{1}{3} g_{ij} K 
\nonumber
\ee
is the extrinsic curvature tensor. $K$ and $A_{ij}$ are the trace and 
trace-free parts of $K_{ij}$, respectively. Covariant differentiation 
with respect to $g_{ij}$ is denoted by ${\nabla}_i$. 
Spacetime indices will be denoted by Greek letters and spatial indices 
by Latin letters.  

The systematic solution of the constraint equations is due to Lichnerowicz,
Choquet-Bruhat, York and others \cite{YP}. This procedure involves freely 
specifying a metric $\tilde{g}_{ij}$, a traceless extrinsic 
curvature $\tilde{A}_{ij}$ and the trace $\tilde{K}$. It then introduces a
conformal factor $\phi$, and a scaling rule to determine the physical $g_{ij}$
and $K_{ij}$:
\begin{eqnarray}
g_{ij} = \phi^4 \tilde{g}_{ij},~~
A_{ij} = \phi^{-2} \tilde{A}_{ij},~~
K = \tilde{K}
\nonumber
\end{eqnarray}
as functions of the coordinates (defined both in the physical and in the trial
conformal space). Solution consists of simultaneously determining $\phi$ (Eq.
(\ref{HC}) becomes an elliptic equation for $\phi$) and correcting the
longitudinal part of $A_{ij}$. Typically, one introduces a vector potential
$w^i$ to accomplish this latter task and the problem consists of four 
coupled elliptic equations for $\phi$ and $w^i$ \cite{MHS}.

\section{Kerr-Schild Slicing}

Our work is based on descriptions of black holes in ingoing 
Eddington-Finkelstein (iEF) coordinates. This choice
is motivated by the fact that, in these coordinates, surfaces of
constant time ``penetrate'' the event horizon.
Foliations that penetrate the horizon facilitate the excision of
the singularity from the computational domain.  The essence of black
hole excision is the removal of the singularity while preserving the
integrity of the spacetime accessible to observers outside the black
hole. As originally suggested by Unruh \cite{thornburg}, this is only
possible if the excised region is fully contained within the event
horizon, thus the need to have access to the interior of the black
holes.

Other work, such as that of Br\"{u}gmann \cite{PWS}
for the generic 3-dimensional code, does not use excision, and instead
solves $K=0$ initial data in a conformally flat background
\cite{Punctures,Cook}. Here, we keep $K$ non-zero to maintain close 
similarities to the analytically known single Kerr-Schild black hole.

The 4-dimensional form of the Kerr-Schild spacetime \cite{KS}:

\begin{eqnarray}
ds^2 = - dt^2 +dx^2 + dy^2 + dz^2 +2 H(x^{\alpha}) (l_{\lambda}
dx^{\lambda})^2
\label{interval}
\end{eqnarray}
describes isolated single black holes.  Here the scalar function $H$ has
a known form, and ${ l_{\lambda}}$ is an ingoing null vector congruence
associated with the solution. For instance for the Schwarzschild
solution, $ H(x^{\alpha}) = M/r$, and  $l_{\lambda}= (1;x^i/r)$, {\it i.e.} an
inward pointing null vector with unit spatial part.  From the fact that
Eq. (\ref{interval}) is an exact solution to the Einstein equations, one 
can write the 3-metric, and the momenta $K_{ij}$ associated with Eq.
(\ref{interval}).

Initial data setting for multiple black hole spacetimes using the 
method described by Matzner, Huq, and Shoemaker (MHS) \cite{MHS} 
begins by specifying a conformal spatial metric which is a straightforward
superposition of two Kerr-Schild single hole (spatial) metrics:

\begin{eqnarray}
\tilde{g}_{ij}dx^i dx^j =
\delta_{ij}dx^i dx^j + 2~{}_1H(x^{\alpha}) ({}_1l_{j}dx^{j})^2  \nonumber\\
	+ 2~{}_2H(x^{\alpha}) ({}_2l_{j}dx^{j})^2 ~~.
\label{gapprox}
\end{eqnarray}
The fields marked with the pre-index $1$ ($2$) correspond to an isolated 
black hole with specific angular momentum $\bf{a_1}$ ($\bf{a_2}$) 
and boosted with velocity $\bf{v_1}$ ($\bf{v_2}$).
The superposition of the conformal momenta is defined as follows:
The extrinsic curvature for a single hole (say hole 1) 
\begin{eqnarray}
	{}_{1}K_{ij} = 
	({}_{1}\partial_j \beta_i + 
         {}_{1}\partial_i \beta_j 
	- 2~{}_{1}\Gamma^{k}_{ij}
	{}_{1}\beta_{k} - {}_{1} \partial_t g_{ij}) / (2~{}_{1}\alpha)~~,
\nonumber
\end{eqnarray}
is converted to a mixed-index object,
\begin{eqnarray}
	{}_{1}K_i^{~j} = {}_{1}g^{nj} {}_{1}K_{in}~~.
\nonumber
\end{eqnarray}
The trace of $\tilde{K}$ is calculated as the sum of the corresponding 
traces:
\begin{eqnarray}
	 \tilde{K} = {}_{1}K_i^{~i}+{}_{2}K_i^{~i}~~,
\nonumber
\end{eqnarray}
and the transverse-traceless part of the extrinsic curvature 
$\tilde{A}_i^{~j}$ as
\begin{eqnarray}
	\tilde{A}_i^{~j} = {}_{1}K_i^{~j}+{}_{2}K_i^{~j}- \frac{1}{3} 
	\delta_i^{~j} \tilde{K} ~~.
\label{Aapprox}
\end{eqnarray}

MHS propose the use the metric and extrinsic curvature so 
defined as a {\it conformal} metric and extrinsic curvature to solve the 
coupled elliptic system \cite{Symm}.

\section{Approximate solutions}

Here we take a different approach. After all, when two black holes are widely
separated we expect almost-linear superposition to hold, so the data setting
as specified above, {\it without} the elliptic solution for $\phi$ and $w^i$,
should lead to only small errors in the widely separated case. We show here
that even for interestingly close separation scenarios, the superposition
errors are small, 
and in fact both the $l_1$ and $l_\infty$ norms 
 are smaller than those of the truncation error (the discretization error 
in the calculation) in simulations at currently accessible computational 
resolutions. In all cases here we present the results for head-on collisions 
(which are simpler to display) but very similar results are found at similar 
separations for non head-on data.

\begin{figure}[htb]
\begin{center}
\hskip 2.0 cm
\vskip 1.0 cm
\mbox{\psfig{figure=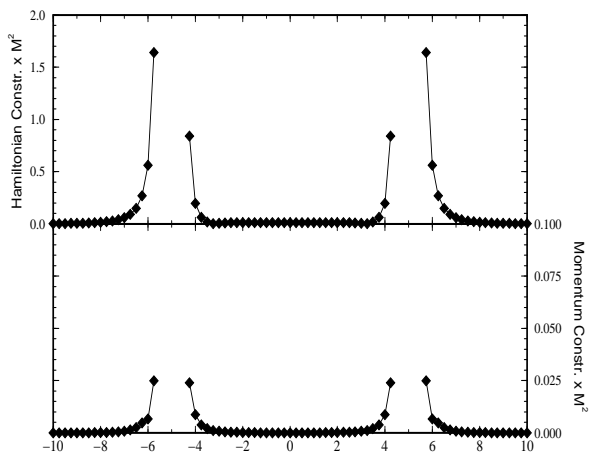,width=6 cm,height=6 cm}}
\end{center}
\vskip 0.5 cm
\caption{Hamiltonian and momentum constraint
residuals for a head-on collision of two equal-mass $M$ black holes. 
The vertical axis shows the residuals of the Hamiltonian
constraint $|\tilde H|$ and the $x$ component of the momentum constraint 
$|\tilde M_x|$ defined by Eqs. (\ref{Hres},\ref{Mres}). The black holes are 
separated $10M$ and boosted toward each other at $0.5c$. Both black holes 
have specific angular momentum $0.5M$ pointing perpendicular to the colliding 
direction. The momentum residual corresponds to the component parallel 
to the colliding axis.}
\label{hc_view}
\end{figure}

Figure \ref{hc_view} shows plots of the norms of the residuals 
of the Hamiltonian and momentum constraints for a head-on collision 
using initial data $\lbrace \tilde g_{ij},\, \tilde A_{ij}\rbrace$ provided 
by Eq. (\ref{gapprox}) and Eq. (\ref{Aapprox}), respectively. The parameters 
of the data are 
$M_1 = M_2 = M = 1$, $|{\bf a_1}| = |{\bf a_2}| = a = 0.5M$ along the $z$ axis, 
and the holes are boosted against each other in the $x$ direction with velocity 
$v=0.5$. The residuals are defined as the absolute value of
\begin{eqnarray}
\tilde H &=& \tilde R+\frac{2}{3} \tilde K^2 - \tilde A_i^{~j} \tilde
A_j^{~i}\label{Hres}~~,\\ 
\tilde M_i &=& \widetilde \nabla_j \tilde A_{i}^{~j} 
- \frac{2}{3} \widetilde\nabla_i \tilde K \label{Mres}~~;\
\label{MCt}
\end{eqnarray}
where again ``$~\tilde{}~$" denotes analytic quantities evaluated using the 
approximate solutions (\ref{gapprox}) and (\ref{Aapprox}). We see from figure 
\ref{hc_view} that the violation of the constraints is primarily confined 
to a region near each hole. This is due to the fact that Eq. (\ref{Hres})
is the sum of terms which scale as $O(M^2/r^4 \times M/d)$, 
where $r$ is the coordinate distance to the singularity and $d$ is the
coordinate separation between holes. For an isolated black hole,
these terms cancel each other out exactly, since the Kerr-Schild metric
is an exact solution of the constraint equations. However, 
that is not the case for the spacetime metric and extrinsic curvatures 
provided by Eqs. (\ref{gapprox}) and (\ref{Aapprox}). The presence
of the perturbation of the second hole in the vicinity of the first hole 
destroys the balance between these terms, causing the right hand side 
of Eq. (\ref{Hres}) to scale as $O(M^2/r^4 \times M/d)$. As in our 
computational approach to evolving the spacetime, we mask the interior of the 
black hole. In this example, we excise the points inside a sphere of radius 
$a+h$ centered at the hole, where the absolute value of the specific angular 
momentum $a$ is also the radius of the Kerr ring-like singularity and $h$ is 
the grid spacing (for figure \ref{hc_view}, $h=M/4$). Hence, figure 
\ref{hc_view} plots precisely the error {\it on the computational domain}.

\begin{figure}[htb]
\begin{center}
\hskip 2.5 cm
\vskip 1.0 cm
\mbox{\psfig{figure=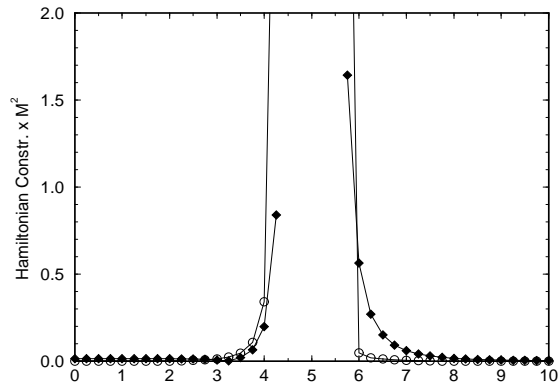,width=6 cm,height=5 cm}}
\end{center}
\vskip 0.0 cm
\caption{Hamiltonian constraint near the rightmost hole for the system of 
figure \ref{hc_view}. The diamonds show the Hamiltonian constraint residuals 
$|\tilde H|$ for the superposition method. The circles are the estimation
$|\bar H|$ of the truncation error present in a second order finite 
difference code ($h=M/4$), defined by Eq. (\ref{Htc}). }
\label{hc_zoom_sup}
\end{figure}

Every finite difference code will have a corresponding truncation error 
associated with discretization. In most of the finite difference codes 
under development, this error scales as $O(h^2)$, second order accuracy; 
therefore, given a second order finite difference discretization of the 
constraints, the truncation errors are obtained from
\begin{eqnarray}
H_{tr} &=& \bar R+\frac{2}{3} \bar K^2 - \bar A_i^{~j} \bar
A_j^{~i}\label{Htrunc}\\
M_{i~tr} &=& \bar \nabla_j \bar A_{i}^{~j}
- \frac{2}{3} \bar\nabla_i \bar K \label{Mtrunc}\,.
\end{eqnarray}
Above, ``$~\bar{}~$" is used to denote finite difference discretization,
and it is understood that these finite difference expressions are to be 
evaluated using the exact, not approximate, solutions of the 
constraints. Of course, these exact solutions are not yet available
\cite{WIP00}. In order to circumvent this problem, we obtain an estimate 
of the truncation errors from the absolute value of
\begin{eqnarray}
\bar H &=& \bar R+\frac{2}{3} \bar K^2 - \bar A_i^{~j} \bar A_j^{~i}
- \tilde H\label{Htc}\\
\bar M_i &=& \bar \nabla_j \bar A_{i}^{~j}
- \frac{2}{3} \bar\nabla_i \bar K - \tilde M_i\label{Mtc}\,
\end{eqnarray}
instead. Although these truncation errors correspond to a modified
form of the constraints, these errors should not be substantially 
different since the structure of the terms involving finite difference 
operators was not changed, only analytic functions ($\tilde H$ and 
$\tilde M_i$) have been added. Notice that for an isolated hole, the 
Kerr-Schild spatial metric and extrinsic curvature satisfy exactly the 
constraint equations ($\tilde H = \tilde M_i = 0$), leaving us with an error 
exclusively related to the truncation error. Figure \ref{hc_zoom_sup} shows 
the residual $\tilde H$ (diamonds) and truncation error $\bar H$ (circles)
in the Hamiltonian constraint in a region around the rightmost hole of figure
\ref{hc_view}. The truncation error $\bar H$ was obtained in a mesh 
with grid spacing $h=M/4$.

\begin{figure}[htb]
\begin{center}
\hskip 2.5 cm
\vskip 1.0 cm
\mbox{\psfig{figure=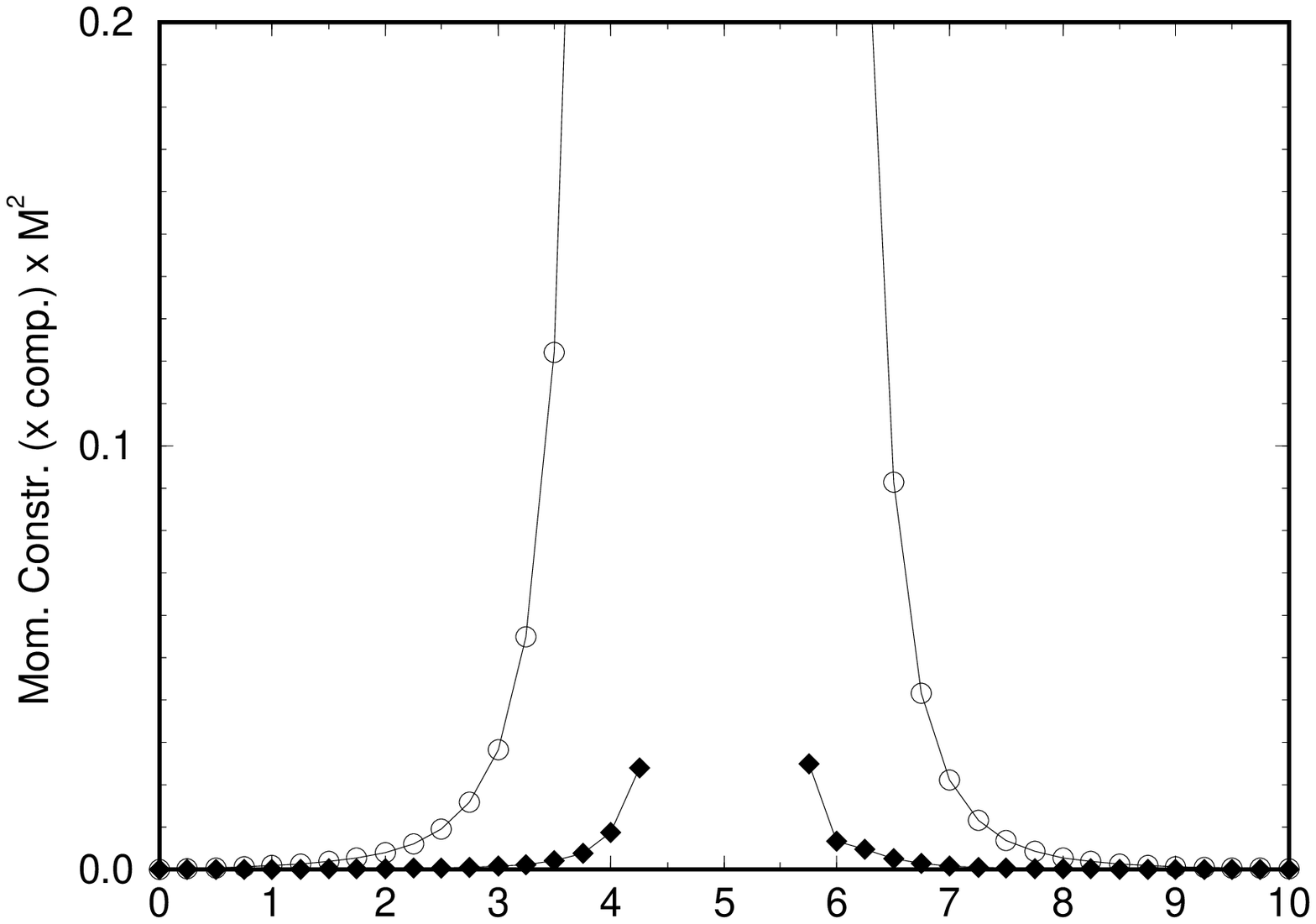,width=6 cm,height=5 cm}}
\end{center}
\vskip 0.0 cm
\caption{Momentum constraint (x component) for the system of figure 
\ref{hc_zoom_sup}. The diamonds show the momentum constraint residuals 
$|\tilde M_x|$ for the superposition method. The circles are the estimation 
$|\bar M_x|$ of the truncation error ($h=M/4$), defined by Eq. (\ref{Mtc}). }
\label{mc_zoom_sup}
\end{figure}

In order to compare the global degree of satisfaction of the initial value 
equations by the methods described here, we calculate the $l_\infty$ (maximum) 
and $l_1$ (average of the absolute value) norms of the Hamiltonian and momentum 
constraints. We do this over the axis joining the black holes from $x=-10M$ to 
$x=10M$, excluding the points inside two segments of length $0.75 M$, centered 
around each hole, as indicated above.
As shown in Table I, lines 1 and 4, both the $l_1$ and the 
$l_{\infty}$ norms of the superposition error are less than 
those of the $M/4$ truncation error. However figure \ref{hc_zoom_sup} 
makes it clear that the truncation error is not uniformly (pointwise) 
larger than the superposition error. We find in practice that at $M/4$, 
superposition yields adequate behavior in the evolution.
Figure \ref{mc_zoom_sup} shows a similar diagram to figure 
\ref{hc_zoom_sup} for the momentum constraint. In this case the 
superposition method can be used to provide data with error pointwise 
uniformly smaller than the truncation error.

\section{Attenuated superposition method}
\label{att_sec}

In order to reduce the residual errors, we propose a variation of the 
superposition method that preserves the simplicity of being analytical. 
Essentially, the method consists of multiplying ``attenuation" functions 
into the recipe of the previous section. The new approximate metric 
$\tilde g_{ij}$, trace $K$ and tensor $\tilde A_i^{~j}$ take the form:

\begin{eqnarray}
\tilde {g}_{ij}^A &=& \delta_{ij}
	    + 2~{}_{1}B(x^k){}_{~1}H(x^k)_{~1}l_{i~1}l_{j}\nonumber\\
	& & + 2~{}_{2}B(x^k){}_{~2}H(x^k)_{~2}l_{i~2}l_{j}~~,\nonumber\\
\tilde{K}^A &=& {}_{1}B{}_{~1}K_i^{~i}+{}_{2}B{}_{~2}K_i^{~i}~~,\nonumber\\
\tilde{A}_i^{A~j} &=& {}_{1}B{}_{~1}K_i^{~j}
	+{}_{2}B{}_{~2}K_i^{~j}
	- \frac{1}{3} \delta_i^{~j} \tilde{K}^A~~.
\nonumber
\end{eqnarray}

The purpose of the attenuation function $B$ is to minimize the effects due 
to a given hole on the neighborhood of the other hole. For instance, the 
attenuation function ${}_{1}B{}$ is unity everywhere except in the vicinity 
of hole-2 where it rapidly vanishes so the metric and extrinsic curvature 
there are effectively that of a single black hole. The attenuation 
functions with this property can be constructed in a number of different 
ways. An example of an attenuation function is
\be
{}_{1}B= 1-e^{-r^4/\sigma^4}~,
\label{FD_att}
\ee
where
\ba
r^{2} &=& \frac{1}{2}(\rho^{2} - a^{2}) +
\sqrt{\frac{1}{4}(\rho^{2} - a^{2})^2 + a^{2}z^{2}}~~,\\
\rho &=& \sqrt{{}_{2}\gamma^2(x-{}_{2}x)^{2} + (y-{}_{2}y)^{2}
+ (z-{}_{2}z)^{2}}~.
\ea
Here ${}_{2}x^i$, $a$ and ${}_{2}\gamma$ denote the coordinate location, 
specific angular momentum and boost factor of hole-2, respectively.  
$\sigma$ represents a free parameter of the attenuation function. 
An expression similar to (\ref{FD_att}) for ${}_2B$ is obtained by 
reversing the labels.

The attenuation function (\ref{FD_att}) was chosen amongst a few different 
types for its simplicity and better performance. Since the constraints 
involve second order derivatives, it is important to pick attenuation 
functions that vanish up to second order derivatives, as in (\ref{FD_att}), 
so ``pure" single black hole solutions can be obtained in the neighborhood 
of each hole.

\section{Results}

\begin{figure}[htb]
\begin{center}
\hskip 2.5 cm
\vskip 1.0 cm
\mbox{\psfig{figure=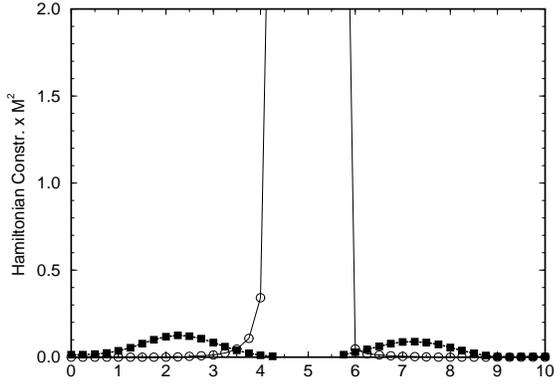,width=6 cm,height=5 cm}}
\end{center}
\vskip 0.0 cm
\caption{Hamiltonian constraint for the system of figure \ref{hc_view}. 
The squares present the Hamiltonian constraint residuals $|\tilde H|$ for the 
attenuation method. The circles are the estimation $|\bar H|$ of the 
truncation error present in a second order finite difference code ($h=M/4$), 
defined by Eq. (\ref{Htc}).}
\label{hc_zoom_att}
\end{figure}

\begin{figure}[htb]
\begin{center}
\hskip 2.5 cm
\vskip 1.0 cm
\mbox{\psfig{figure=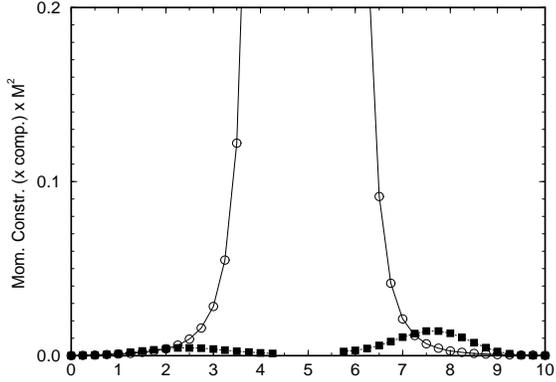,width=6 cm,height=5 cm}}
\end{center}
\vskip 0.0 cm
\caption{Momentum constraint (x component) for the system of figure 
\ref{hc_zoom_att}. The squares show the momentum constraint residuals 
$|\tilde M_x|$ for the attenuation method. The circles are the estimation 
$|\bar M_x|$ of the truncation error ($h=M/4$), defined by Eq. (\ref{Mtc}). }
\label{mc_zoom_att}
\end{figure}

Figures \ref{hc_zoom_sup} and \ref{hc_zoom_att}
(\ref{mc_zoom_sup} and \ref{mc_zoom_att}) show the Hamiltonian (momentum)
constraint for a region around the rightmost hole of figure \ref{hc_view}.
Figure \ref{hc_zoom_sup} shows a comparison between the truncation error 
for $h=M/4$ (empty circles) and the superposition initial data (full diamonds),
while figure \ref{hc_zoom_att} shows the same for attenuated initial data
(full squares). 

First we note in figure \ref{hc_zoom_att} that in some areas (near $x=2M$, for
instance) the violation of the constraints for these attenuated data is greater 
than that corresponding to truncation error. However, even in these areas
the violation is smooth and small in absolute value, as opposed to the unbounded
behavior shown by the superposition data near the singularity, and 
the divergent behavior present at the singularity has 
disappeared with the use of attenuated data. This is due to the fact that, 
having cancelled the influence of the presence of the second hole, the fields 
$\tilde{g}^A_{ij}$ and $\tilde{K}^A_{ij}$ become the fields corresponding to 
an isolated black hole at the location of the singularity.

The results are presented in Table \ref{table_norms}. The momentum 
constraint $l_1$ norm represents the average of the three components 
of Eq. (\ref{MCt}) and the $l_\infty$ the maximum value of them. These 
values are compared to the norms of the truncation error in the constraints 
caused by the finite difference stencils for grid spacings $h=M/4$, $M/8$,
and $M/16$. These results show that with these norms, the superposition initial 
data violate the Hamiltonian and momentum constraints below the truncation 
error corresponding to grid spacings as fine as $h=M/4$, while the attenuated 
data extend this range to at least $h=M/8$. Note also that, for both the 
superposition and the attenuated method, the momentum constraint equations 
are satisfied well within the truncation error for grid spacings as fine 
as $h=M/16$. Finally Table \ref{table_norms} (rows 2, 3) shows
the approximate second order convergence between resolutions $h=M/8$ and
$h=M/16$. The convergence rate is {\it approximately} second order because
even at $h=M/16$, finite $h$ effects affect significantly the convergence.

\section{Conclusions}

The initial value problem for black hole binary systems is currently 
approached by solving numerically a set of elliptic equations derived 
from the Hamiltonian and momentum constraint. Due to the presence of 
singularities in the fields, these equations are complex and difficult 
to handle.

In this article we presented analytical methods that can
provide initial value data for black hole binary systems. While
these solutions are only approximate, they satisfy the Hamiltonian and 
momentum constraint equations to the level of accuracy 
(evaluated with the $l_1$ and $l_\infty$ norms) of the truncation 
error present in finite difference codes for a given range of grid 
resolutions. Because attenuated data are so much smoother than the
superposition, and because the principal difficulties in computational 
simulations arise from sharp gradients and divergent values, which the 
attenuation method eliminates,we are confident that at least for the particular 
example of a head-on collision of two equal-mass black holes at a separation 
of $10M$, thess data could be used in evolutionary codes with grid spacings as
fine as $h=M/8$.

The analytical nature of these methods makes them simpler to implement 
than the numerical approach. Furthermore, a more comprehensive study of 
the attenuation functions may extended the range of resolutions for which 
this approximation is valid.

The attenuation method offers the advantage that it cancels the divergent
behavior of the constraints near the central singularity.
The suppression of contributions near the black holes (individual holes 
have exactly zero residuals) should simplify the exact 
problem, since this method provides exact inner boundary conditions
for the elliptic solver.

\section{Acknowledgement}

This work was supported by NSF grant PHY9800722 and PHY9800725 to 
the University of Texas at Austin and PHY9800973 and PHY9800970 to the
Pennsylvania State University.

\begin{table}
\caption{Truncation errors (first three rows) and residuals
(last two rows) of the Hamiltonian and momentum constraints.}
\begin{tabular}{lcccc}
\ Method & \multicolumn{2}{c}{Ham. Constr.} & \multicolumn{2}{c}{Mom.
Constr.} \\
  & $l_\infty$ & $l_1$ & $l_\infty$ & $l_1$ \\
 \tableline
&&&&\\
$h=M/4$             &   $4.240$ & $0.251$ & $20.20$ & $0.605$\\
$h=M/8$             &   $0.757$ & $0.022$ & $3.098$ & $0.065$\\
$h=M/16$            &   $0.198$ & $0.004$ & $0.742$ & $0.012$\\
without attenuation &   $1.643$ & $0.097$ & $0.235$ & $0.003$\\
with attenuation    &   $0.126$ & $0.049$ & $0.041$ & $0.005$\\
\end{tabular}
\label{table_norms}
\end{table}


\begin{thebibliography}{99}
\bibitem{non-norm} For this article we work with the constraint functions
instead of their normalized counterparts. The normalized constraints
provide the degree of violation compared to the magnitude of the terms in
each equation and can be misleading when these terms have a
relatively small absolute value.
\end{thebibliography}
\end{document}